\documentclass[a4paper,fleqn,usenatbib]{mnras}

\usepackage[T1]{fontenc}
\usepackage{ae,aecompl}

\usepackage{graphicx}	% Including figure files
\usepackage{amsmath}	% Advanced maths commands
\usepackage{amssymb}	% Extra maths symbols
\usepackage{caption}

\title[Martian atmospheric loss]{Did high-energy astrophysical sources contribute to Martian atmospheric loss?}

\author[Dimitra Atri]{
Dimitra Atri,$^{1}$\thanks{E-mail: dimitra@bmsis.org}
\\
$^{1}$Blue Marble Space Institute of Science, 1001 4th Ave, Suite 3201, Seattle, WA 98154\\
}

\date{Accepted XXX. Received YYY; in original form ZZZ}

\pubyear{2016}

\begin{document}
\label{firstpage}
\pagerange{\pageref{firstpage}--\pageref{lastpage}}
\maketitle

% Abstract of the paper
\begin{abstract}
Mars is believed to have had a substantial atmosphere in the past. Atmospheric loss led to depressurization and cooling, and is thought to be the primary driving force responsible for the loss of liquid water from its surface. Recently, MAVEN observations have provided new insight into the physics of atmospheric loss induced by ICMEs and solar wind interacting with the Martian atmosphere. In addition to solar radiation, it is likely that its atmosphere has been exposed to radiation bursts from high-energy astrophysical sources which become highly probable on timescales of $\sim$Gy and beyond. These sources are capable of significantly enhancing the rates of photoionization and charged particle-induced ionization in the upper atmosphere.  We use Monte Carlo simulations to model the interaction of charged particles and photons from astrophysical sources in the upper Martian atmosphere and discuss its implications on atmospheric loss. Our calculations suggest that the passage of the solar system though dense interstellar clouds is the most significant contributor to atmospheric loss among the sources considered here.
\end{abstract}
%Here, we explore the possibility of damage from Galactic Gamma Ray Bursts, nearby supernovae, encounter with dense interstellar clouds and extreme solar events.
% Select between one and six entries from the list of approved keywords.
% Don't make up new ones.

%%%%%%%%%%%%%%%%%%%%%%%%%%%%%%%%%%%%%%%%%%%%%%%%%%

%%%%%%%%%%%%%%%%% BODY OF PAPER %%%%%%%%%%%%%%%%%%

\section{Introduction}

The robotic exploration of Mars has uncovered a wealth of geological features such as gullies and channels indicating flowing water on its surface in the past. Given its distance from the Sun, liquid water on its surface can only be maintained with a significant atmosphere dominated by greenhouse gases. The present atmospheric surface pressure is about 6 mb on an average and is composed of 95.32\% CO$_2$, 2.7\% N$_2$ and 1.6\% Ar by volume. This raises a big question, how did Mars lose its atmosphere? One of the main differences between the Earth and Mars is the lack of a significant magnetic field and smaller surface gravity on Mars. A magnetic field provides shielding from incoming charged particles and also prevents escape of outgoing ions. A combination of small escape velocity and the lack of magnetic shielding enables easier escape of ions from its upper atmosphere. The Sun, being the primary source of incoming charged particles, generates solar wind, solar flares and coronal mass ejections (CMEs). The spectrum of ionizing radiation from the Sun spans from EUV, X-rays, gamma ray photons to keV - MeV protons, extending up to $\sim$10 GeV in extreme cases. The ionosphere is created by photoionization from solar EUV. 

Recently, Mars Atmosphere and Volatile EvolutioN or MAVEN, which is currently operating in the Martian orbit, provided us with key measurements of ion escape rates in the upper atmosphere. The major escaping ion species are O$^+$, O${_2}^{+}$, H$^+$, and the minor ones include CO${_2}^{+}$ and H${_2}^{+}$. The EUV and X-ray fluxes causes photodissociation, photoionization, ion-electron pair production which leads to photoelectron escape. The primary escape mechanisms include photochemistry, recombination of ions and electrons, charge exchange between ions and photochemical escape of atoms \citep{jakosky2015initial}. A fraction of the energy is lost in the form of heat, airglow and transport processes such as thermal conduction and molecular diffusion. The electrons take away a large fraction of the kinetic energy since positive ions are heavy and possess less recoil energy \citep{schunk2009ionospheres}. The other mechanisms include ion sputtering produced by ions impacting the upper atmosphere, plasma energization and solar wind forcing and ion pickup by solar wind electric field. MAVEN also discovered permanent plume like features on mars with escaping ions \citep{dong2015strong}. On longer timescales, it is highly likely that the Martian atmosphere was exposed to intense radiation bursts from a number of astrophysical sources. Energetic photons and charged particles can significantly increase ionization rates in the upper atmosphere and aid in particle escape based on mechanisms described above. 

On timescales of 100 - 1000 years, solar events such as the well-known Carrington event of 1859 and the 774/775 AD radiation burst (based on $^{14}$C spike in tree rings) become important \citep{miyake2012signature,melott2012causes}. $^{60}$Fe deposits found on the ocean floor (multiple locations around the globe) as well as on the Moon, indicate the occurrence of multiple nearby supernovae within $\sim$100 pc from the solar system \citep{knie2004f,binns2016observation,breitschwerdt2016locations,fimiani2016interstellar,wallner2016recent}. Studies have also indicated 7-8 instances of enhanced flux of Anomalous Cosmic Rays during 0.25 Gyr due to the passage of solar system through dense interstellar clouds \citep{pavlov2005passing}. Calculations also suggest  a significant probability of a Galactic GRB beaming towards a planet in the past 1 Gyr \citep{melott2004did}. 

The main objective of this manuscript is to calculate the energy deposition rate on the top layer of the martian atmosphere from these sources and subsequently discuss their role in affecting the the atmospheric depletion rates. In some cases we will use data from particular sources, but we emphasize that our goal is to estimate the fluence of multiple events in Gyr timescales. This approach is also reasonable because of the temporal variability in the emissions from these sources.

\section{Fluence estimates from astrophysical sources}
\subsection{Nearby Supernovae}
The idea that nearby supernovae can inject energy and damage planetary atmospheres was given by Ruderman (1974). Recent reports provide evidence of a 2.2 Myr old supernova at a distance between 60-130 pc based on measurements of $^{60}$Fe deposition on the ocean floor \citep{binns2016observation,breitschwerdt2016locations,wallner2016recent} and on the Moon \citep{fimiani2016interstellar}. They also report the possibility of 16 events in the past 13 My within 100 pc in the Local Bubble. They estimate the frequency of such nearby explosions, with at least one event  occurring every 2-4 My based on SN rate of $\sim$ 2 events/century \citep{wallner2016recent}. On longer timescales of $\sim$ Gyr, one can assume that a large number of such events would have deposited energy in the Martian atmosphere. Let us now calculate the distance - fluence relation for a typical event. We base our calculations on the work done by \cite{gehrels2003ozone}, where they computed the atmospheric effects of nearby supernovae on Earth. They estimated a rate of $\sim$ 1.5 events per Gyr within a distance of 8 pc. Since the gamma ray spectrum of SN 1987A was used for calculations, it had to be rescaled due to the unusual nature of of its progenitor, a blue supergiant. 
\begin{center}
$\frac{dN}{dE}=1.7 \times 10^{-3} \Big(\frac{E}{1 \,MeV}\Big)^{-1.2}$ cm$^{-2}$\,s$^{-1}$\,MeV$^{-1}$
\end{center}
Rescaled gamma ray energy of 1.8$\times$$10^{47}$erg at a distance of 100 pc gives a fluence of 9.4$\times$$10^{16}$eV\,cm$^{-2}$\,sr$^{-1}$ on top of the Martian atmosphere. It should be noted that the total energy released in a core-collapse event is $\sim$ $10^{51}$ erg. Only a fraction of that energy is released in form of gamma rays and accelerated particles. Recent observations of IC 443 and W44 have suggested that the energy released in cosmic rays is between 1 to 10\% of the total energy or $10^{49}$ to $10^{50}$ erg \citep{ackermann2013detection}.
\begin{center}
$\frac{dN}{dp} = A\,p^{-s_1}\Big[1+\big(\frac{p}{p_{br}}\big)^{\frac{s_2 - s_1}{\beta}}\Big]^{-\beta}$
\end{center}
Where p is the momentum, A is the normalization constant, $\beta$ describes the smoothness of the break in the power law and was set to 0.1, $p_{br}$ is the location of the spectrum break, s$_{1}$ and s$_{2}$ are indices below and above the break. This gives a fluence of 5.2$\times$$10^{18}$eV\,cm$^{-2}$\,sr$^{-1}$ - 5.2$\times$$10^{19}$eV\,cm$^{-2}$\,sr$^{-1}$at 100 pc. These estimates represent the upper limit of fluence because we did not consider cosmic ray diffusion from nearby sources. SN generated particles in the $\sim$ 10 GeV-PeV range will deposit only a small fraction of the energy in the upper atmosphere and most of their energy will be deposited deeper in the atmosphere and cause other effects, not relevant here. 

\subsection{Galactic GRBs}
Based on observations, the local occurrence rate of GRBs pointing towards a particular target is estimated to be $\sim$ 6 $\times$ 10$^{-11}$\,y$^{-1}$\,kpc$^{-2}$ \citep{melott2004did,thomas2005gamma}. They go further and estimate the probability of impact to be 18\% at 1 kpc, 55\% at 2 kpc and 70\% at 4 kpc distance. These numbers suggest a significant probability of a Galactic GRB beaming towards Mars on Gyr timescales. We assume the jets to be purely leptonic based on the non-detection of hadronic component coincident with GRBs by the IceCube experiment. Gamma rays emitted in GRB jets are in the keV - MeV energy range, and are capable of ionizing and depositing a significant amount of energy in the upper atmosphere. \\

$\frac{dN(E)}{dE}$ = A$\Big(\frac{E}{100\,keV}\Big)^\alpha\exp\Big(\frac{-E}{E_0}\Big),$ $ E \leq (\alpha - \beta)E_0$  
\vspace{0.2cm}
\hspace{1.5cm}= A$\Big[\frac{(\alpha - \beta)E_0}{100 \,keV}\Big]^{\alpha - \beta} \exp(\beta - \alpha) \Big(\frac{E}{100 \,keV}\Big)^\beta$, \\$E \geq (\alpha - \beta)E_0$.

The fluence of a ``typical" Galactic GRB at 6 kpc estimated by \cite{melott2004did} is 100\,kJ\,m$^{-2}$, which equates to 6.24$\times$$10^{19}$eV\,cm$^{-2}$ incident on the top of Martian atmosphere. Most events are accompanied by an afterglow in X-rays with a peak at around 25 keV, but the emitted energy is about 1\% of the prompt emission. We will include the afterglow in our calculations because as we show later, they also contribute significantly to ionization in the upper atmosphere. Typical GRBs last a few seconds and could potentially increase the ionization rate significantly for a short period.

\subsection{Encounter with Interstellar Clouds}
The study of atmospheric effects of enhanced ACR (Anomalous Cosmic Rays) flux resulting from passage of the solar system through dense interstellar cloud coupled with magnetic field reversals was first done by \cite{pavlov2005passing}. Dense molecular clouds have neutral atoms and molecules, some of whom are ionized and accelerated within the solar system. The enhanced flux typically lasts for $\sim$ 1 Myr. The density of incident material is upwards of 100 H atoms cm$^{-3}$ and can go up to 330 H atoms cm$^{-3}$. The interaction takes place with the velocity of about 20 km\,s$^{-1}$. They also conclude that 7-8 such episodes might have occurred in the last 0.25 Gyr. Mars lacks a significant magnetic field and one can envisage that such events would have significant implications for the Martian atmosphere. Also, there would be multiple CMEs during the $\sim$1 Myr window that might further accelerate the escape rate. Under this scenario, the plume-like structures on Mars are likely going to change. We follow the method prescribed in \cite{pavlov2005passing} to compute the spectrum of incident charged particles. The ACR spectrum depends on two factors, the local interstellar density - n(H) and distance from the termination shock - r$\prime$$_{TS}$, and can be expressed in form of an equation shown below. \\

F(E,r) = F$_{0}$(E,r$^{0}$$_{TS}$) $\times \frac{n(H_{cloud})}{n(H_{PLIM})} \times \exp(\frac{u(r - r\prime_{TS})}{K})$ \\

Where K (radial diffusion coefficient) = {\it Lv}/3, {\it L} is the particle mean free path, {\it v} is the particle velocity, {\it u} is solar wind speed, {\it r} is heliospheric distance to mars, PLIM is present local interstellar medium and r$^{0}$$_{TS}$ present position of the termination shock. With a moderately dense interstellar cloud, n(H) = 150 atoms cm$^{-3}$, the termination shock moves closer with r$\prime$$_{TS}$ = 2 A.U. and the fluence of incident particles is of the order of $\sim$10$^{12}$ eV\,cm$^{-2}$s$^{-1}$. %from its present value of 100 A.U. The flux on top of the atmosphere comes out to 2.4$\times$10$^8$ eV\,cm$^{-2}$s$^{-1}$ considering only protons.

\subsection{Extreme Solar Events}
Solar events include solar flares, coronal mass ejections and possibly more extreme events such as superflares, observed on sun-type stars. MAVEN observations \citep{jakosky2015initial} report typical solar wind proton density of 1.8 cm$^{-3}$, alpha of 0.1 cm$^{-3}$ with a speed of 505 km\,s$^{-1}$. During interaction with an ICME (Interplanetary Coronal Mass Ejection) the measurements change to $10^4$ particles cm$^{-3}$ with a speed of 820 km\,s$^{-1}$. These ICMEs last for a few hours and are thought to be the main sources of atmospheric loss. Since MAVEN has already measured the impact of ICMEs on the martian atmosphere, and studied the effects it had on the ion escape rates, we will not repeat the analysis of such ``regular" events. Instead, we focus on less frequent but more intense events. The emissions from solar events is distributed in two channels, photons in the EUV and X-ray band, and energetic protons mostly in the keV - MeV energy range. However, some ``hard" solar proton events are accompanied by protons of energies beyond 1 GeV and in some cases found to go up to 10 GeV. The most intense flares, classified as X-class flares are major eruptions with X-ray flux (1-8 angstrom) of $> 10^{-4}$ W\,m$^{-2}$. Multiple events with omnidirectional fluence $> 10^9$ cm$^{-2}$ (on earth) for energies $>$ 30 MeV have been reported to occur within a solar cycle \citep{smart2006two}. The largest recorded event, known as the Carrington event occurred 1859, with estimated energy of $\sim$ $2\times10^{33}$ erg and event integrated proton fluence of $ \sim 3\times 10^{10}$ cm$^{-2}$ on earth. It caused disruptions to the telegraph network and auroras were observed around the globe. Recently, 774/5 AD and 993/4 AD events were discovered based on spikes in $^{14}$C concentration, along with $^{10}$Be and $^{36}$Cl spikes found on multiple locations around the globe. The estimated fluence of the 774/5 AD is estimated to be $\sim$ $9\times10^{33}$ erg \citep{cliver2014solar}.

We now move on to other intense flares observed on G type stars similar to the sun, also known as superflares. These flares are characterized by fluence of typically six order of magnitude larger than the observed solar flares. Earlier, the based on the occurrence rate of such events, it was thought of as an exception. In a study, nine Superflares were spotted on solar type stars with omnidirectional fluence between $10^{33}$ to $10^{38}$ erg \citep{schaefer2000superflares}. Recent work by \cite{karoff2016observational} analyzed 5648 stars including 48 superflare stars. Based on their analysis, they concluded that the possibility of such flares ($> 2\times10^{34}$ erg) occurring on the sun cannot be ruled out.

\section{Numerical Modeling}
We use the GEANT4 simulation package to model the propagation of high-energy protons and photons in the Martian atmosphere \citep{agostinelli2003geant4}. It is developed at CERN and considered a gold standard in radiation propagation calculations.The software is used worldwide and has been validated with a variety of experiments over the years. It has the capability to simulate the interactions of very low energy photons of a few eV all the way up to Galactic Cosmic Rays, or particles with energies of 10 GeV and beyond. The energy of the primary particle is used up in a variety of processes such as scattering, heat, ionization etc. GEANT4 takes all processes into account and gives the energy deposition rates in the atmosphere. The package has also been used for similar studies by others \citep{pavlov2005passing,pavlov2011dense}. We obtain spectra of various sources from section 2 and use them to input in our code. To begin with, we assume the atmosphere to be 100\% CO$_2$ for simplicity. We consider the current composition for calculations later. Since the size of the Martian atmosphere has evolved considerably in the Gyr period, we focus our efforts in understanding the energy deposition of photons and particles as a function of column density instead of altitude. This approach is also reasonable from a physical standpoint, because as we note below, the energy deposition rate depends on two factors, the energy of the incident particle and the column density of the absorber. The mean energy loss rate, $\frac{dE}{dx}$ can be calculated using the well-known Bethe-Bloch equation \citep{dormancosmic}. The calculations done here can be easily adopted to run on sophisticated atmospheric models by converting column density to height. Since mechanisms relevant here occur in the upper atmosphere, we take into account only the top 1 g\,cm$^{-2}$ of the atmosphere and discard all radiation penetrating at lower levels of higher column density. This would equate to an altitude above 75 km for a primordial atmosphere of 1033 g\,cm$^{-2}$ column density (same as present earth) and 0.385 bar atmospheric pressure described in detail in \cite{dartnell2008computer}. For each run we generated $10^9$ events to reduce statistical errors. Background rates were obtained from the HELIOSARES model \footnote{\url{https://lasp.colorado.edu/maven/sdc/public/pages/models.html}}. The final numbers were renormalized based on fluence obtained in section 2. 

\section{Results}
Figure 1 shows the results of the interaction of protons with energies from $10^6$ eV to 10$^{15}$ eV in form of the total energy deposited in the top 1 g\,cm$^{-2}$ Martian atmosphere. Most of the energy of primaries with kinetic energies up to 100 MeV is deposited in the 1 g\,cm$^{-2}$ atmosphere and the interactions are all electromagnetic. At kinetic energies above 290 MeV, which is the pion production threshold, a significant fraction of the particle's total energy is diverted in producing hadrons. Production of particles such as pions at higher energies takes away the energy from electromagnetic interactions \citep{atri2011modeling}. The figure also shows that at energies of 1 GeV and above, particles deposit only a very small fraction of their energy in the top layer. The energy deposition increases gradually with primary energy. Higher energy primaries deposit bulk of their energy deeper in the atmosphere \citep{atri2010lookup}. Figure 2 shows the results of the interaction of photons with energies from 10 eV to 10$^{8}$ eV in form of total energy deposited in top 1 g\,cm$^{-2}$ martian atmosphere. Figure 2 shows the energy deposition of photons from 10 eV to 10$^9$ eV. It can be seen that photons up to 10 keV deposit all their energy in the 1 g\,cm$^{-2}$ atmosphere. At 100 keV and above, most of the energy is deposited in the lower layers and is therefore discarded. The energy deposition slowly rises with primary energies of 100 keV and beyond.

\begin{figure}
\centering % \begin{center}/\end{center} takes some additional vertical space
\includegraphics[width=0.5\textwidth]{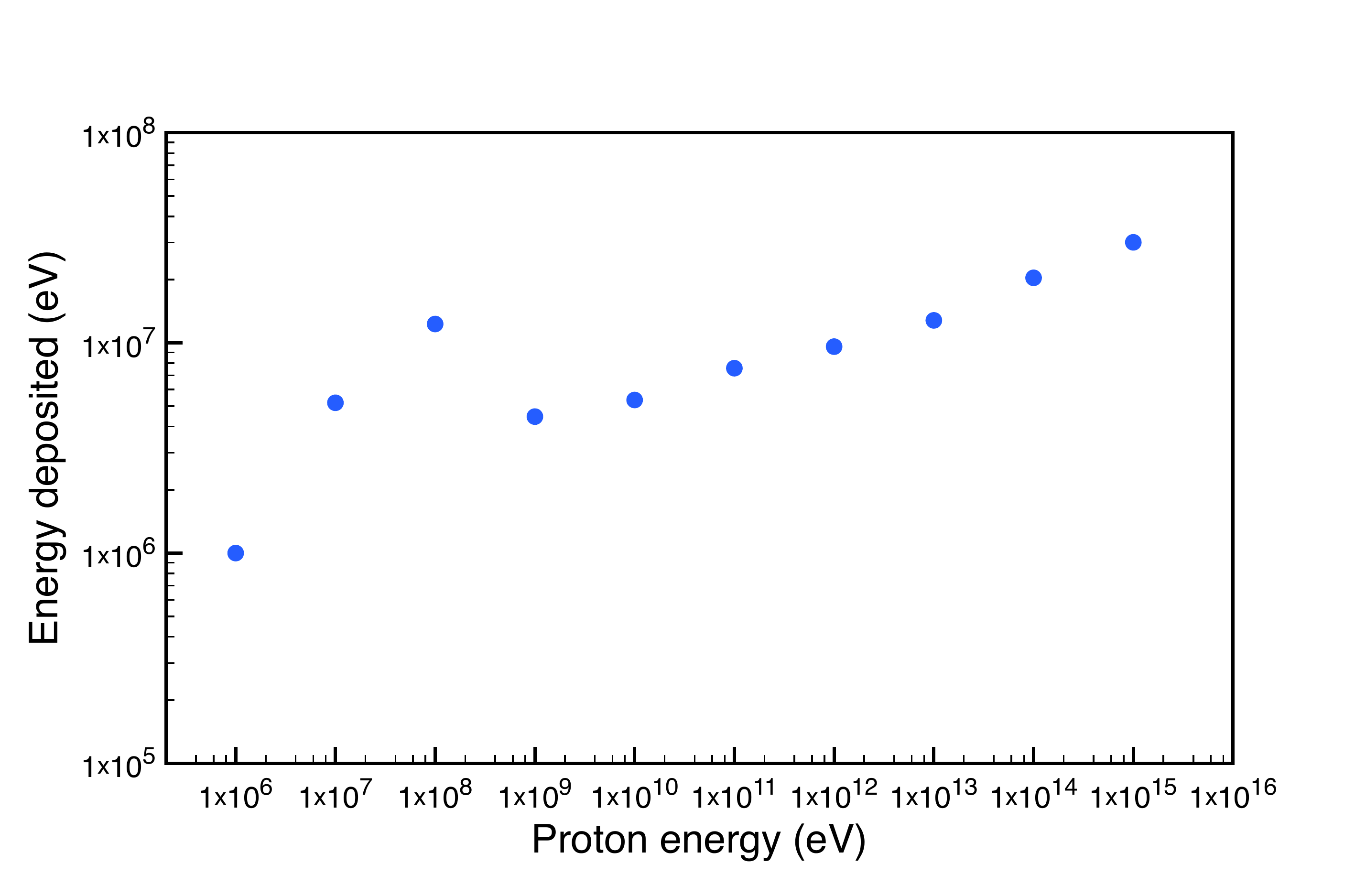}
\caption{\label{fig:i} Energy deposited by incident protons in the top 1 g\,cm$^{-2}$ Martian atmosphere. The kinetic energy of protons is represented on the horizontal axis.}
\end{figure}

\begin{figure}
\centering % \begin{center}/\end{center} takes some additional vertical space
\includegraphics[width=0.5\textwidth]{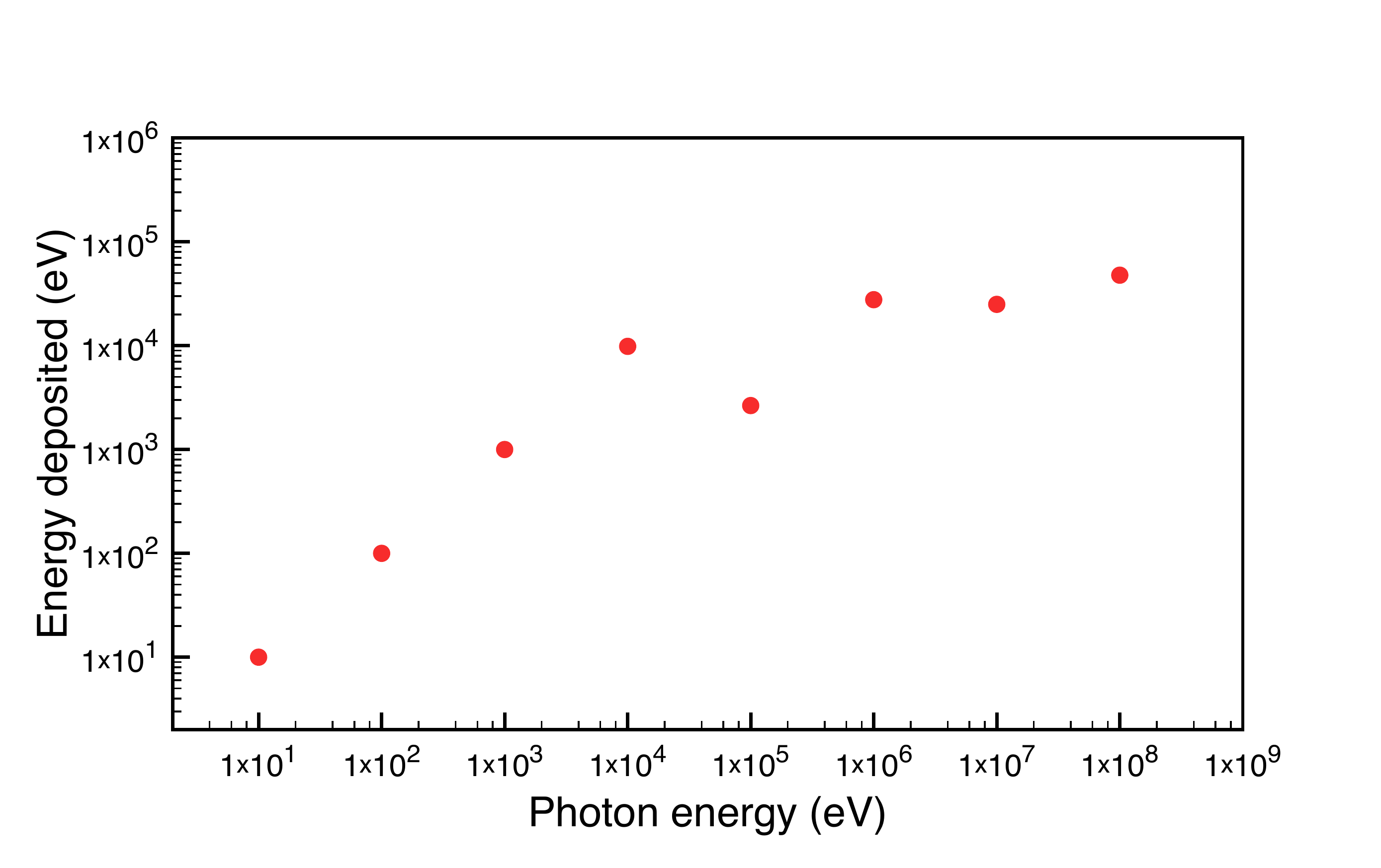}
\caption{\label{fig:i}Energy deposited by incident photons in the top 1 g\,cm$^{-2}$ Martian atmosphere. The energy of incident photons is on the horizontal axis.}
\end{figure}

All the energy deposition calculations described here are for the top 1 g\,cm$^{-2}$ atmosphere. For nearby supernova we treat gamma and particle emissions separately. For gamma emission, we use the renormalized spectrum of SN 1987A following the earlier work of \cite{gehrels2003ozone}. We found the gamma ray-induced energy deposition to be 3.74$\times10^{15}$ eV\,cm$^{-2}$\,sr$^{-1}$ at 100 pc. Assuming that the radiation is spread over 300 days, the deposition rate turns out to be 9$\times10^8$ eV\,cm$^{-2}$s$^{-1}$. For comparison, the Superthermal and Thermal Ion Composition (STATIC) instrument aboard MAVEN measured the energy of planetary ions during peak flux to be $\sim 10^8$ eV\,cm$^{-2}$$^{-1}$s$^{-1}$ \citep{dong2015strong}. For charged particles from nearby supernova, we took the spectral parameter values from \cite{ackermann2013detection} for W44 (measured by Fermi-LAT) assuming that 10\% of the total energy was emitted in cosmic rays. We obtain 1.6$\times10^{17}$eV\,cm$^{-2}$ spread over 100 years giving the rate of 2.6$\times10^8$ eV\,cm$^{-2}$s$^{-1}$ at 100 pc. For the Milky Way GRB scenario, we take Band parameter values: E$_0$ = 187.5 keV, $\alpha$ = -0.8, $\beta$ = -2.3, spread over 1 keV - 10 MeV range \citep{thomas2005gamma}. We obtain 3.96$\times10^{18}$ eV\,cm$^{-2}$ deposited in the top 1 g\,cm$^{-2}$ of the atmosphere. The spectral information for the Interstellar cloud scenario was obtained from \cite{cummings2002composition} following the method in \cite{pavlov2005passing,pavlov2011dense} giving the total fluence of 8.8$\times10^{11}$ eV\,cm$^{-2}$s$^{-1}$. For modeling extreme solar events, estimating the spectrum is difficult because most of the events occurred before the satellite age. We selected the spectrum of the 20 January 2005 event here, because even though the event was not large by any standards, it was very well studied by a number of groups using both satellite and ground based measurements. We rescaled its total energy to the Carrington event ($2\times10^{33}$ erg) and took the Martian distance into account. The energy deposition rate was found to be 3$\times10^{11}$ eV\,cm$^{-2}$s$^{-1}$ assuming the event was spread over 24 hours.

%\begin{center}
%    \begin{tabular}{| l | l |}
    %\hline
   
    %\bf{Source} & \bf{Ion pairs} (cm$^{-2}$s$^{-1}$)\\ \hline
    %SN ($\gamma$) at 100 pc & 2.8$\times10^7$\\ \hline
    %SN (particle) at 100 pc & 3.8$\times10^7$ \\ \hline
    %GRB & 1.2$\times10^{16}$\\ \hline
    %Interstellar Cloud & 2.7$\times10^{10}$\\ \hline
    %1859 Carrington event & 9.2$\times10^9$\\ \hline
    %Superflare & 9.2$\times10^{10}$ \\ \hline
    
    %\end{tabular}
%\end{center}

The ion pair generation energy {\it w} of CO$_2$ is 33 eV ion pair$^{-1}$. The primary mechanism at play is photochemical escape, for example: 
  
\begin{center}
CO$_2 \rightarrow$ CO$_2^+$ + e$^-$\\ CO$_2^+$ + O $\rightarrow$ O$_2^+$ + CO\\ O$^+$ + CO$_2$ $\rightarrow$ O$_2^+$ + CO etc. 
\end{center}
%Dissociative recombination:\begin{center} O$_2^+$ + e$^-$ $\rightarrow$ O + O. \end{center}
\begin{figure}
\centering % \begin{center}/\end{center} takes some additional vertical space
\includegraphics[width=0.5\textwidth]{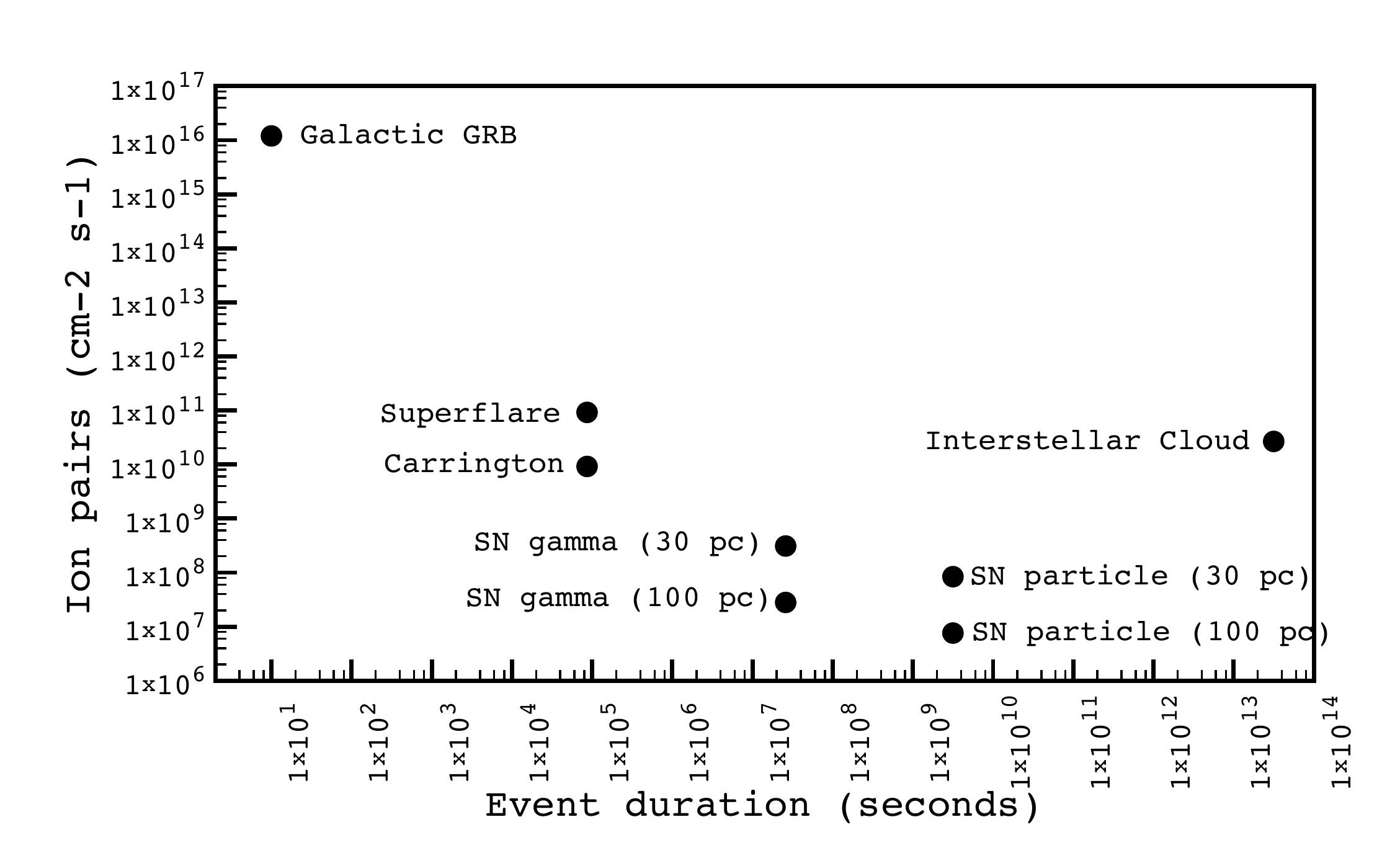}
\caption{\label{fig:i}Production rate of ion pairs in the top 1 g\,cm$^{-2}$ Martian atmosphere for all the events. Event duration is shown on the horizontal axis.}
\end{figure}
A complete description of all the reactions and associated energetics can be found in standard textbooks such as \cite{schunk2009ionospheres}. We divide the deposited energy obtained from simulations by the ion pair generation energy to calculate ion pairs for each scenario. Our results are spread over a wide range of ion production rates, 1.2$\times10^{16}$ ion pairs cm$^{-2}$s$^{-1}$ from a Galactic GRB to  2.8$\times10^7$ ion pairs cm$^{-2}$s$^{-1}$ from gamma rays of a supernova at 100 pc. Figure 3 shows the ion production rates of different events along with their durations. Particles can escape the Martian atmosphere with just 0.125 eV/amu energy. With charge exchange, the fast ion becomes neutral and escapes, and the slower ion gets trapped in the magnetic field. Fast ions can directly knock the atoms out of the atmosphere, this process is known as sputtering. Ions also escape by gaining energy from ambipolar diffusion and by thermal escape according to Jeans equation. A Galactic GRB would directly deposit energy, produce ions (1.2$\times10^{16}$ ion pairs cm$^{-2}$s$^{-1}$) and transfer energy to kinetic energy of ions. The event is short lived but the magnitude of ionization and energy transfer is large. Thermal expansion due to a large energy deposition is possible in this scenario. In the nearby supernova case, the gamma-ray and cosmic ray enhancements are $\sim$1 and $\sim$100 years long respectively. In addition to ion escape by large energy transfer, a fraction of the ions will also be picked up by solar wind and ICMEs during the period. Encounter with dense interstellar clouds would result in a $\sim$1-Myr-long enhancement in ionization rates ($\sim10^{10}$ ion pairs cm$^{-2}$s$^{-1}$). Extreme solar events are short-lived but as seen in figure 3, can do significant damage and are relatively more frequent. 

The efficiency of the individual escape processes is difficult to estimate, especially with varying atmospheric and astrophysical parameters on Gyr timescales. Overall escape rates are also difficult to estimate for long duration events because as we discuss later, the probability of a combination of events affecting the escape rate becomes high. Therefore, we make some reasonable assumptions to estimate the total atmospheric loss in all these scenarios. We assume that the escaping ions are equally divided into O$^+$ and O${_2}^{+}$, and calculate the upper limit of the escape rate assuming 100\% efficiency. The upper limit of atmospheric loss in each scenario is shown in table 1. For comparison, recent MAVEN observations report a background ion escape rate of $\sim$ 2$\times$10$^{24}$ s$^{-1}$ \citep{brain2015spatial} and 3.34$\times10^{25}$ s$^{-1}$ or 1.27 kg/s during an ICME \citep{jakosky2015initial}. With a conservative estimate that the efficiency of escape is 10\%, the ion escape rate ranges from $\sim$10$^{32}$ ions for protons from supernova at 100 pc to $\sim$10$^{41}$ ions for the interstellar cloud scenario. We also carried out simulations with the present atmospheric composition \citep{mahaffy2015structure} with higher concentration of oxygen in the upper atmosphere. The results indicate higher escape rates as shown in the table. The total mass of the Martian atmosphere is $\sim$ 2.5 $\times$ 10$^{16}$ kg. The encounter with dense interstellar cloud seems to be the most significant scenario with an upper limit of loss of about 0.5\% of the total present Martian atmosphere. We further explore this scenario with a 0.1 bar (268 g\,cm$^{-2}$) and 0.01 bar (27 g\,cm$^{-2}$) atmosphere \citep{dartnell2008computer}. For altitude above 75 km, for CO${_2}$ atmosphere, the loss was calculated to be 1.0$\times10^{14}$, 6.9$\times10^{13}$, 2.4$\times10^{13}$ kg, and for present composition, 1.3$\times10^{14}$, 8.9$\times10^{13}$, 3.1$\times10^{13}$ kg for 0.385 bar, 0.1 bar and 0.01 bar respectively. With thinner atmospheres, the loss reduces to 0.35\% with 0.1 bar and 0.13\% with 0.01 bar atmosphere. 

\begin{center}
\captionof{table}{Upper limit of atmospheric loss in kg} 

    \begin{tabular}{| c | c | c |}
    \hline
   
    \bf{Source} & CO$_2$ atmosphere & Present composition\\ \hline
       1859 Carrington event & 9.5$\times10^4$ & 1.3$\times10^5$\\ \hline
       Superflare & 9.5$\times10^{5}$ & 1.3$\times10^6$\\ \hline
       SN ($\gamma$) at 30 pc & 9.6$\times10^5$ & 1.3$\times10^6$\\ \hline
       Galactic GRB & 1.4$\times10^{7}$ & 2$\times10^{7}$\\ \hline
       SN (particle) at 30 pc & 3.2$\times10^7$ & 3.6$\times10^7$\\ \hline
	Interstellar Cloud & 1.0$\times10^{14}$ & 1.3$\times10^{14}$\\ \hline

    \end{tabular}
\end{center}
%\begin{center}
%\captionof{table}{Upper limit of atmospheric loss in the interstellar cloud scenario} 

%    \begin{tabular}{| c | c | c | c |}
 %   \hline
   
%    \bf{Total loss (kg)} & 0.385 bar & 0.1 bar & 0.01 bar\\ \hline
 %    CO${_2}$ atmosphere & 1.0$\times10^{14}$ & 6.9$\times10^{13}$ & 2.4$\times10^{13}$  \\ \hline
%     Present composition & 1.3$\times10^{14}$ & 8.9$\times10^{13}$ & 3.1$\times10^{13}$  \\ \hline
%    \end{tabular}
%\end{center}
\section{Discussion}
Planetary atmospheres play a major role in maintaining habitable conditions on a planet. Studies of atmospheric degradation mechanisms have brought attention to the importance of planetary magnetism and surface gravity in preventing atmospheric loss. Over the years, a number of probes on Mars have provided us with valuable data to understand these processes. We now have excellent observations of frequent ICMEs and solar wind playing an important role in atmospheric depletion in the upper Martian atmosphere. We extend these results to model the possibility of atmospheric depletion from high-energy astrophysical sources. We have adopted a simple approach to estimate the upper limit of escape of O$^+$ and O${_2}^{+}$ ions. For all long-duration events, a combination with other events can accelerate the escape process. Possible mechanisms include direct ionization of the upper atmosphere providing ions enough energy to escape Martian gravity. For example, ICMEs during the supernova cosmic ray or interstellar cloud scenarios. In ancient past, the Sun was highly active and produced stronger winds and the magnitude and frequency of extreme events was higher. This would increase the rate of ion escape considerably. In addition to mechanisms described above, atmospheric loss is possible through direct sputtering (dense incident protons) in case of passage through dense interstellar clouds. Although, all astrophysical sources discussed in this manuscript contribute to atmospheric loss, the encounter with dense molecular cloud appears to be the most effective in atmospheric depletion due to long duration of the event. A loss of $\sim$ 10$^{14}$ kg becomes highly significant given the frequency of such encounters, which is about 7-8 times in the last 0.25 Gyr and $\sim$135 times in the last 4.5 Gyr. We emphasize that these are preliminary results because we do not consider the escape efficiency, thermal and other escape processes into account and further modeling of individual escape processes can be done with sophisticated MHD models by using our results as inputs. Such an effort would help in determining the efficiency of individual escape processes and in better understanding the role of high-energy astrophysical sources in planetary atmospheric loss. 

%We propose that a combination of events, where in addition to direct ionization, ions get picked up by ICMEs and solar wind would accelerate the depletion process. This is relevant especially for long-duration events and in ancient scenarios when the Sun was highly active. 

\section*{Acknowledgements}

The author would like to thank A. L. Melott, J. Haqq-Misra and the anonymous reviewer for their helpful comments. Atmospheric data was obtained from the MAVEN Science Data center at the suggestion of P. Mahaffy. Computations were carried out with GEANT4 (geant4.cern.ch) and ROOT (root.cern.ch) packages using the Extreme Science and Engineering Discovery Environment, supported by the National Science Foundation grant number ACI-1053575. 

%%%%%%%%%%%%%%%%%%%%%%%%%%%%%%%%%%%%%%%%%%%%%%%%%%

%%%%%%%%%%%%%%%%%%%% REFERENCES %%%%%%%%%%%%%%%%%%

% The best way to enter references is to use BibTeX:

\bibliographystyle{mnras}
\bibliography{mars} % if your bibtex file is called example.bib

% Alternatively you could enter them by hand, like this:
% This method is tedious and prone to error if you have lots of references

%%%%%%%%%%%%%%%%%%%%%%%%%%%%%%%%%%%%%%%%%%%%%%%%%%

%%%%%%%%%%%%%%%%% APPENDICES %%%%%%%%%%%%%%%%%%%%%

%%%%%%%%%%%%%%%%%%%%%%%%%%%%%%%%%%%%%%%%%%%%%%%%%%

% Don't change these lines
%\bsp	% typesetting comment
\label{lastpage}
\end{document}